# AN ANNOTATION-BASED APPROACH TO SUPPORT DESIGN COMMUNICATION

**Onur Hisarciklilar and Jean-François Boujut**
G-SCOP Laboratory, INPG Grenoble University

**ABSTRACT**
The aim of this paper is to propose an approach based on the concept of annotation for supporting design communication. In this paper, we describe a co-operative design case study where we analyse some annotation practices, mainly focused on design minutes recorded during project reviews. We point out specific requirements concerning annotation needs. Based on these requirements, we propose an annotation model, inspired from the Speech Act Theory (SAT) to support communication in a 3D digital environment. We define two types of annotations in the engineering design context, locutionary and illocutionary annotations. The annotations we describe in this paper are materialised by a set of digital artefacts, which have a semantic dimension allowing express/record elements of technical justifications, traces of contradictory debates, etc.

In this paper, we first clarify the semantic annotation concept, and we define general properties of annotations in the engineering design context, and the role of annotations in different design project situations. After the description of the case study, where we observe and analyse annotations usage during the design reviews and minute making, the last section is dedicated to present our approach. We then describe the SAT concept, and define the concept of annotation acts. We conclude with a description of basic annotation functionalities that are actually implemented in a software, based on our approach.

## 1 INTRODUCTION

Engineering design processes are becoming more and more complex. Since many years, successful design methodologies based on the decomposition of these processes into sub-processes or tasks have been developed in order to deal with complex design situations. Companies have implemented design procedures in order to control the quality of the design process and eventually asses the quality of the product. However, today's market environment requires more than complex procedures in order to asses the quality of the product. New organizations based on concurrent engineering principles involve co-operative work of an increasingly high number of stakeholders from different fields of expertise during the design process.

Improving communication between actors having different conceptual words, representations and tools, is a great challenge in today's organisations. Co-operative artefacts (drawings, sketches, CAD models, simulation results, etc.) play an important role in design communication. According to Vinck and Jeantet [1, 2], they are used as intermediary objects, i.e. they are related to the action itself (i.e. the product), and they are means for co-ordinating designers' activity. Star [3] stresses their role as boundary objects, allowing the expression of a shared knowledge between cross-domain actors.

On the other hand, communication between actors involved in these organisations is still suffering from a lack of efficiency. In fact, actors are spending a large part of their time seeking, organising, modifying and translating information, often unrelated to their own personal disciplines [4]. In spite of the multitude of computer-supported tools (such as CAD, CAM or simulation tools) which aim to support specific issues during the design processes, there is very limited number of tools dedicated to support design communication. Actors often need to be provided with means for developing more systematic co-operation around product representations and more adapted information to the context of use [5].

This paper proposes a Speech Act Theory (SAT) based annotation model to support communication in 3D CAD models. Our annotation model is a set of digital artefacts, which have a semantic dimension that improves the design communication through the elicitation of knowledge related to the context of design.

The paper is divided into seven sections. In the next section, we clarify semantic annotation concept, define general properties of annotations in the engineering design context, and the role of annotations in different design project situations. In the third section, we describe a case study based on our observations in an industrial vehicle company, we observed and analysed annotation practices during the design reviews and minute making. The forth section is dedicated on our approach. We first describe the SAT concept, and define the annotation acts, inspired from that concept. Finally, we propose basic annotation functionalities, based on our approach.

## 2 SEMANTIC ANNOTATIONS

Semantic annotations can be described simply as annotations that are interpretable (or reusable) by a human being in a given context. In contrast, computationally semantic annotations are dedicated to be interpreted by machines. Annotations that are dedicated to human utilization are called cognitively semantic annotations (here we will call them semantic annotation).

In the Semantic Web domain, where annotations are primarily defined in order to be computed by machines, semantic annotations are described as annotations that refer to an ontology. In other words, semantic annotations formally identify concepts and relations between concepts in documents [6]. Semantic Web annotation brings two kinds of benefits to the information systems, enhanced information retrieval and improved interoperability. Information retrieval is improved by the ability to perform searches, which exploit the ontology to make inferences about data from heterogeneous resources. Interoperability is particularly important for organizations, which have large legacy databases, often in different proprietary formats that do not easily interact. In these circumstances, annotations based on a common ontology can provide a common framework for the integration of information from heterogeneous sources.

Semantic annotations dedicated to human utilization, on the other hand, are defined by their goals. The goal of an annotation in this context is the relation between object (information) and the action (the effect of the information). Amongst various works an annotations dedicated to human interaction, Marshall [7], for example, defines six types of annotations in collaborative reading according to their goal: annotations as procedural signals, annotations as place markings and aids to memory, annotations as in situ locations for problem-working, annotations as a record of interpretive activity, annotations as a visible trace of the reader's attention, and annotations as incidental reflections of the material circumstances.

### 2.1 Properties of semantic annotations in engineering design context

Although the exact definition of an annotation is still controversial, we can approach a basic definition of the concept of semantic annotation by listing its properties and particularly by clearly distinguishing it from the concept of document.

Documents are graphical or textual representations (a report, a CAD model, etc.), created to accomplish a task in a given context. Although there may be other documents that can be used complementary to the main document, any document can be interpreted independently from other documents. In our context, a document is mostly a 3D CAD model or any extraction (VRML, etc.) of this model.

In contrast, annotations are attached to a document and can be interpreted only in the context of this document. Although they have this contextual relationship with the document, the goal behind their creation may differ from the goal of the entire document. Annotations are not all the time easy to detect especially when the documents are under construction.

The general properties of annotations in mechanical design context can be summarized as follows [8]:
- An annotation have a different nature from the document on which it is attached to (representing non-geometrical information on a geometrical CAD object in our case).
- The document is the target the annotation refers to.

- The content of an annotation is the information the annotation conveys.
- The anchor of an annotation is the point onto the document, where the annotation is attached.
- The sphere of influence of an annotation is defined by its personal or public status.
- Annotations lifetime is always shorter than the document lifetime.
- The originator and the user of an annotation may be different.

As we have seen, an annotation only makes sense when it is considered with the document it is attached to. The document constitutes therefore the context that makes it possible to understand the information that is conveyed by the annotation.

## 2.2 Role of annotations in collaborative design

Annotations (on sketches, drawings, etc.) have been used since many years in design teams as a mean for communication. The introduction of digital media have radically changed the annotation processes [5], which were traditionally paper based. From the different design situations that we observed, we concluded that the annotations are essentially used across two phases of the design process: asynchronous phase, where the digital artefact is produced, and synchronous phase, where the artefact is collectively evaluated.

An asynchronous situation is defined as a situation where a designer produces a CAD model of an object, or more generally a situation where an individual activity is carried out. In that case, notes can be produced individually in order to establish a list of decisions, remarks, explanations, etc. making reference to a document (e.g. the CAD model). Annotated documents often remain private and can be used for several objectives, such as information indexation, or memorization of the current design situation, etc. Annotations are used in asynchronous situation to represent and capitalize information whose nature is not completely geometrical, such as a manufacturing process, or a type of material, etc.

The other engineering design situation when annotations are often used is a synchronous situation where a collective evaluation of the artefact is carried out. During this activity, intermediary documents are commented and annotated, mostly on a paper base. Today, these meetings are generally mediated by digital representations, and the distant actors communicate through instant messaging and/or video conferencing tools. During these activities, annotations are used mainly as a way to reinforce the oral discourse. Annotations created here are poorly structured and cannot be reinterpreted outside the context where they have been created. Therefore, the majority of annotations created during a design review cannot be reused during another one. All critics and argumentations are oral and nothing remains after the meeting apart from the personal notes taken by the participant of the review.

The design review is a place where solutions are discussed, and the points of view are expressed. Although these evaluations lead sometimes to alter the structure of the product, the solution is very seldom modified during these reviews. A minute is created during the meeting that records the main decisions and is supposed to help the designers during the asynchronous phase. We will see later how this situation leads to misunderstandings.

In conclusion, we consider that annotations play a major role in design coordination and knowledge elicitation in asynchronous phases, and an important cognitive synchronisation role in synchronous phases.

## 3. CASE STUDY

In this section, we will make a description of a co-operative design case, based on our field study in an industrial vehicle company. Our objective here is to expose annotation usage, which can be observed during the project reviews. Annotations are used in this case for recording and communicating the decisions and actions to be performed and also for communicating the context of the decision and some elements of the design rationale (specific constraints for example).

The cross-domain team we will consider here is leaded by an "architect" (the name given to the engineer in charge of the design of a specific aspect of the product). With his high technical level, he coordinates the design activities of an entire sub-system of the truck. He communicates with the designers during asynchronous phases of the project in order to assure the geometrical conformity and coordinates the design reviews. He is responsible of the design solution. The actor called "PMS" (Project Management Support) works with the architect and is in charge of short-term operational management of the project. During the asynchronous phases, he manages and communicates information about the studies in progress (such as deadlines, types of vehicles impacted by each study, etc.). He is also in charge of the design minutes during the design reviews. The "designers" are technical actors who develop solutions in the CAD environment during the asynchronous phases. Another kind of actor, called "script writer", assists the architect by collecting the updated CAD representations and the. Other actors, called "industrials" are experts from different domains (manufacturing, SAS, quality, etc.). They participate to the design reviews in order to evaluate the design solution with regard to their specific constraints.

As we have seen earlier, the asynchronous phase is the period when the designers produce new designs. This activity demands technical knowledge and skill. We point out the fact that the model is developed mainly according to the individual decisions of the designer, on the basis of his own knowledge of the context and decisions taken during previous meetings, i.e. the process relies on the designer's own memory and interpretation.

Although this is an individual activity, the designer needs sometimes to collaborate with the others, especially with the technical actors (the architect, the script writer, or other designers). Communication during these unplanned meetings is made in an unstructured way (face-to-face meetings, telephone calls or email exchanges). They are means to debate or unofficially validate a design solution proposition. Therefore, this is an event where important decisions can be made.

When the model is completed, the script writer integrates this instance into the shared CAD environment. In other words, from that particular moment, the model (the solution) becomes accessible to the other actors, until the next design review.

In the next section, we will show how a SAT based annotation model could improve this aspect of design communication.

### 3.1 Design reviews

The design reviews of our case study were originally dedicated to a validation/rejection process (the procedure defines them as decision points only). However, the stakeholders took the opportunity of these regular meetings to debate on the solution, as there was no other formal design meetings dedicated to that activity in the general design process organisation. This implies that the creative input of a design review is not as secondary as it may seem at a first glance. It is a place where key decisions and their rationale are made explicit and we have had the opportunity to assist to strong discussions in some cases. In our case, as the participants did not have the opportunity to access information about the design decisions before, design reviews became the unique event when participants were able to exchange arguments about the design solution and make new propositions.

### 3.2 Verbal exchanges and design minute

In a design review, first the designer presents the design solution that he produced. It is an oral presentation, when he explains all information that cannot be represented in the CAD model (decisions that he made, key points of the solution, etc). Then, the participants discuss the solution. That is the phase where domain-specific rules are made explicit, and key decisions are made.

The design minute is constructed simultaneously by the PMS during this discussion phase. When a decision is made or an action is planed, he takes a screenshot of the projected screen on his PC, and takes note of the decision or the action by annotating it (see figure 1). The other actors cannot see the PMS's notes during the design review, in other words, they have not the opportunity to share and complement these annotations. Finally, the design minute is a pdf document composed of a series of annotated screenshots. The annotations are of textual nature, anchored to a point on the image by an arrow. After the review, this document is stored on a shared database, and remains accessible to all participants.

We noticed that the CAD model used to visualise the solution is the only co-operation artefact shared during the design review. And the minute is the only shared document after the design review and are composed of static annotated screenshots.

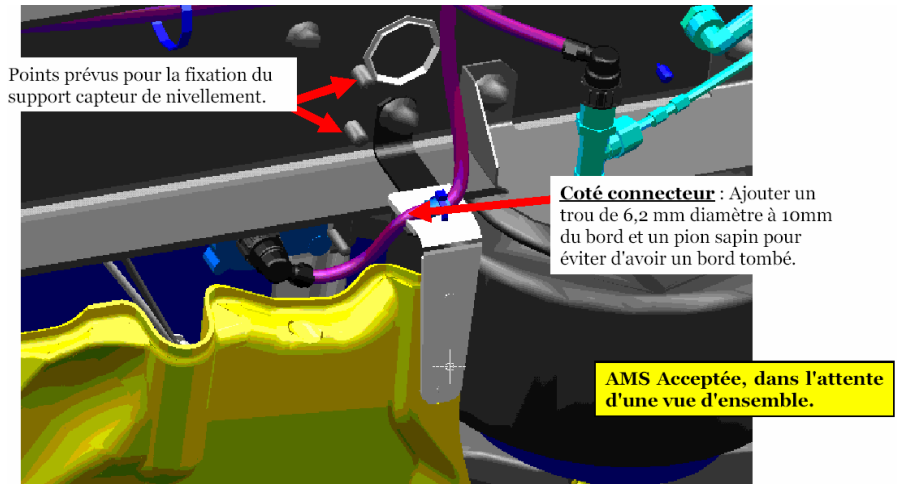

*Figure 1: An annotated screenshot in a design minute*

### 3.3 Observations about annotation usage in the case study

In our case, the annotated CAD model screenshots (the design minute in our case) are produced in order to record the decisions during synchronous phases in a more visual form that better corresponds to the designers practice than a text based minute, and to share them during the asynchronous phase. The design minutes are used by the designers as a memory in order to produce a new version of the CAD model, the design minute is then a means to transmit information from the synchronous phase to the asynchronous one.

We observed many limitations of this practice during the design processes. Firstly, the design minute and annotations are disconnected from the 3D CAD model, which prevents participants from having the possibility to directly annotate the 3D CAD objects, and visualise them at the same time they visualise the CAD object. This also prevents participants from visualising the annotations in a multi-view environment, which makes it difficult to exactly locate the annotations, to visualise the whole environment where the annotation has been put, and sometimes to understand the information that the annotation conveys on a specific part of the model. This situation also discourages the participants from using the annotated design minutes.

In addition to that, there is no annotation structure that allows participants to make free technical discussions around the CAD model. In our case, annotations are created by the PMS during the design review, in order to keep trace of the decisions. A structured annotation tool, on the other hand, could offer the participants the possibility to argument on the solutions, elicit domain-specific information during the design reviews and all along the design process [9]. This can allow the extension of the argumentation phases and enhance the awareness of the participants concerning the design situation.

In the next section, we will describe an annotation model based on the Speech Act Theory (SAT). After the first part of the section where we describe the SAT, we will define the annotation acts. Then, we will illustrate our model on an annotation tool demonstrator.

## 4. TOWARD A STRUCTURED ANNOTATION MODEL

### 4.1 From speech acts to annotation acts

In this section, we describe a methodology to structure annotations. Our approach is based on the speech act theory (SAT), which began roughly with the publication of Austin's How to Do Things with Words [10], and formalized by Searle [11].

The central idea of the speech act theory is that when communicating, people do not just utter propositions; they also perform illocutionary acts, such as requesting, stating, and so on. In other words, the speaker may express different attitudes toward the same propositional content (utterance, or locutionary act, see Table 1). Every speech act consists of an illocutionary force, applied to an utterance. This is represented by the logical relationship, $F(P)$, where P is an utterance, and F an illocutionary force. Moore [12] summarizes this hypothesis by claiming that the logical operator of every utterance (everything we could possibly say) is not Boolean (true or false), nor temporal: it is an illocutionary force. Covington [13] adds that this operator is never vacuous; that is, $F(P) \neq P$. For example, even when stating a fact, you are making a statement, not just voicing a fact.

*Table 1: Various illocutionary acts with similar utterances*

| *Illocution* | *Sentence* |
|---|---|
| Statement | The cat is on the mat |
| Question | Is the cat on the mat? |
| Command | Put the cat on the mat |
| Polite request | Could you put the cat on the mat, please? |
| Promise | I promise that I will put the cat on the mat |
| Offer | I'll put the cat on the mat if you like |

Illocution is distinct from the meaning of a message. In illocution point of view, 'what time is it?' and 'how old are you?' has no difference – they are both questions. The difference between these two appears when the propositional content is concerned.

Illocutionary force is also distinct from the perlocutionary effect of a speech act. The perlocution of an utterance is what it actually accomplishes, such as persuading, informing, etc. Despite the fact that the illocutionary and perlocutionary are closely related, they can be easily distinguished. For example, by requesting you to put the cat on the mat, I may end up either making you to do it or not. In other words, the speaker controls the illocution, but attempts to control the perlocution.

The last and maybe the most important point for our purposes is that the same sentence can contain various illocutionary forces. For example, 'I will finish this work today' can either be a statement, an offer, a promise, or a threat, depending on the context of use.

Effective communication requires accurate recognition of speech acts [13]. Human language requires to elaborate inference in order for the hearer to identify speech acts. In our annotation model, we want to offer the necessary elements to allow interlocutors to express and recognize them.

### 4.2 Annotation acts

Annotating is a natural way for expressing design constraints and providing intermediary representations that support communication between the designers, as observed in [8]. Their function is to represent the various points of view, specific to each expert, and to provide the members of team with the means to take part in and to support discussions concerning these differences in such a way that a shared understanding may be achieved.

On the other hand, our observations show that every annotation expressed in a given context has a specific purpose. Expressing this purpose and its recognition by all actors involved in a design communication is important to achieve an effective communication, and to reach a shared understanding. In that sense, we consider annotations by making an analogy with speech acts. By following the hypothesis of Covington, we consider that effective communication via annotations requires accurate recognition of annotation acts, and propose two types: locutionary and illocutionary annotation acts.

A locutionary annotation act in design context is any information that an actor means to express by an annotation, in other words it is the utterance conveyed by an annotation. When cross-domain communication is concerned, actors need for example to express design constraints, actions to take, or decisions made.

An illocutionary annotation act, on the other hand, is the intention of the actor by putting an annotation on a CAD object (for example, to clarify a solution, to propose a solution, to identify a problem, to evaluate a solution, etc.). A same locutionary annotation act may be put for different purposes in a given context, for example, I can express a constraint to clarify a solution, but also to identify a problem. The illocutionary annotation acts gives in that sense to annotation its conversational dimension, thus have a crucial role to the accurate recognition of the message conveyed by the annotation.

### 4.3 Illustration through the case study

The locutionary and illocutionary annotation acts are implicitly conveyed in our case, in the design minute annotations. Figure 2a and Figure 2b illustrate two annotated CAD screenshots, made after the solution about the tubes circuit of a truck's exhaust heating system had been discussed during a design review. The annotation in Figure 2a, "interference at the exhaust tubes level", were put in that case to identify a problem (interference) in a particular point of the tubes circuit. In other words, the utterance (the locutionary annotation act) "interference at the exhaust tubes level" were put in this example with the intention to (the illocutionary annotation act) "identify a problem" (negative evaluation).

As for the annotation in Figure 2b conveys an utterance (locutionary annotation act), "Idem for the tire suspension, move the tubes of 40mm (or more), in order to avoid the interference and to keep a minimum tolerance of 30mm", in order to "propose a solution" (illocutionary act) to a particular problem.

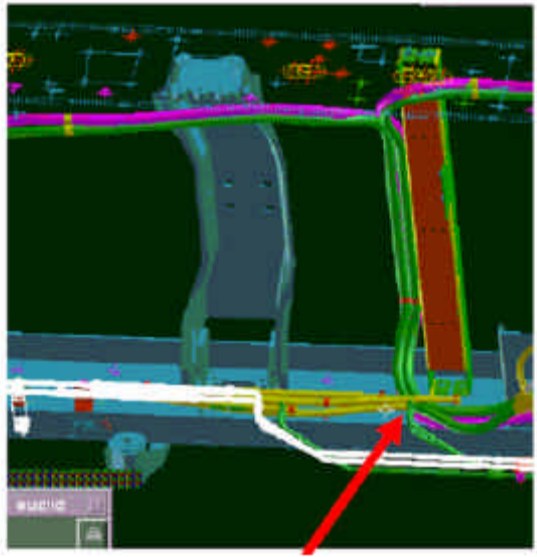 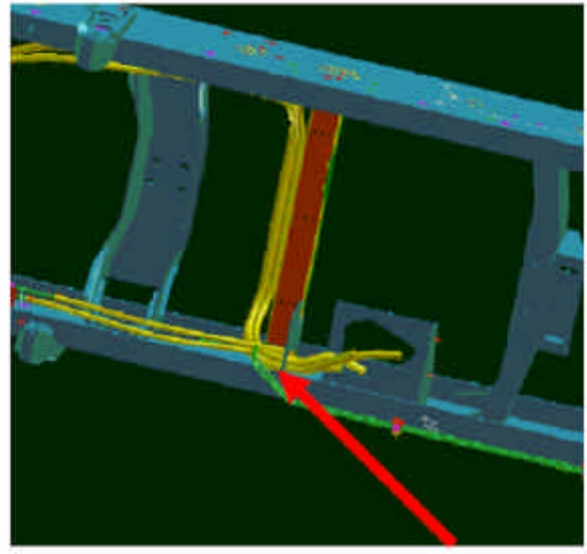

*Figure 2a (left): A problem identification annotation: "interference at the exhaust tubes level" Figure 2b (right): A solution proposition annotation: "Idem for the tire suspension, move the tubes of 40mm (or more), in order to avoid the interference and to keep a minimum tolerance of 30mm"*

## 5.  A SHORT STATE OF THE ART ON 3D GRAPHICAL ANNOTATION TOOLS OR FUNCTIONALITIES

At this point of the paper, it would be interesting to make a state of art of some classical tools offering annotation functionalities, in order to evaluate them regarding our need.

Today, most of the 3D CAD modelling software tools provide designers with some annotation functionalities. However, these annotations are just simple pointers, not having a structure to support design communication in an effective way. On the other hand, some other commercial tools, dedicated to support design collaboration offer more sophisticated annotation functionalities. These tools may be divided in two categories: PDM (or their extension with PLM) systems and tools to support co-design.

### 5.1   3D viewers on PDM systems

3D viewers are tools often provided within design software packages such as CAD and/or PDM environments. It is a class of software that allows distant collaborators to collaborate around 3D models. These programs allow the users to visualize digital mock-ups stored in the PDM during the design development.

This class of software is also designed to support design discussions during design reviews, in synchronous conference mode, where participants of the discussion can annotate the shared model. It can also be used in asynchronous mode, if the user annotates a document that he created himself, or that he received from another user. In both cases, textual annotations and technical annotations (as dimension or distance indication, tolerance, etc.) can be made freely.

While this type of environments is particularly interesting for synchronous technical discussions around 3D models, there is no real annotation structure, which allows the participants to convey, reuse or interpret complex messages.

### 5.2 Annotation tools to support co-design

#### 5.2.1 Online shared models, collaborative modelling

This particular type of tool allows participants to make online design reviews, where participants can share and discuss 3D models. They often have no annotation functionalities that could allow participants to have synchronous communication on the 3D models, but they additionally allow to modify the structure of the product by remotely controlling the CAD application itself.

On the other hand, participants can annotate 2D screenshots of the model, in order to record the minute of the design review. These 2D annotations do not have any structure in order to support a discussion, and allow reuse, or the interpretation in another context.

#### 5.2.2 Asynchronous communication around 2D and 3D drawings

Another type of tool consists of 2D/3D model visualisation tools to support asynchronous design communication offering advanced visualisation functionalities.

These tools allow the user to annotate the model with 3D geometrical symbols placed on the geometry, with a text associated. The file created, which contains the model and the annotations are then transmitted to other collaborators. The annotations are recognized by the tool when the receiver visualises the model, and allow him to accept or refuse the annotations. But the usage of annotations is limited, since the 3D geometrical symbols have no specific meaning, and there is no annotation structure, which could allow participants to make discussions.

This short state-of-art illustrates the lack of an effective annotation tool around 3D models. In the next step of this paper, we will illustrate an annotation tool demonstrator to show how our approach can help the design communication to go further.

## 6. SPECIFICATIONS FOR AN ANNOTATION BASED 3D CAD VIEWER

Based on that approach, we propose a first framework of annotation functionalities. Annotations that we describe here are 3D graphical objects, with a text box associated. The annotation structure is then composed of a two level information structure:

- The first level is the geometrical symbol placed onto the geometry whose form characterises the illocutionary level conveyed by the annotation. In other words, the specific form of the annotation represents the intention of the participant who creates the annotation. We defined three basic acts: proposition, clarification, evaluation and validation. The proposition, evaluation and validation acts represent a solution proposition, a solution evaluation and a solution validation, respectively; while the clarification annotation can represent either a solution clarification or a problem clarification.
- The second level is the forum-based text box attached to the annotation, which conveys the locutionary level of the annotation act. The reason why the locutionary annotation is conveyed by a text is the multitude of the possible locutionary acts that may be expressed. Mostly they are either a design constraint, an action to perform, or a decision made. The participants can edit and modify the content by adding their own remarks, propositions, etc.

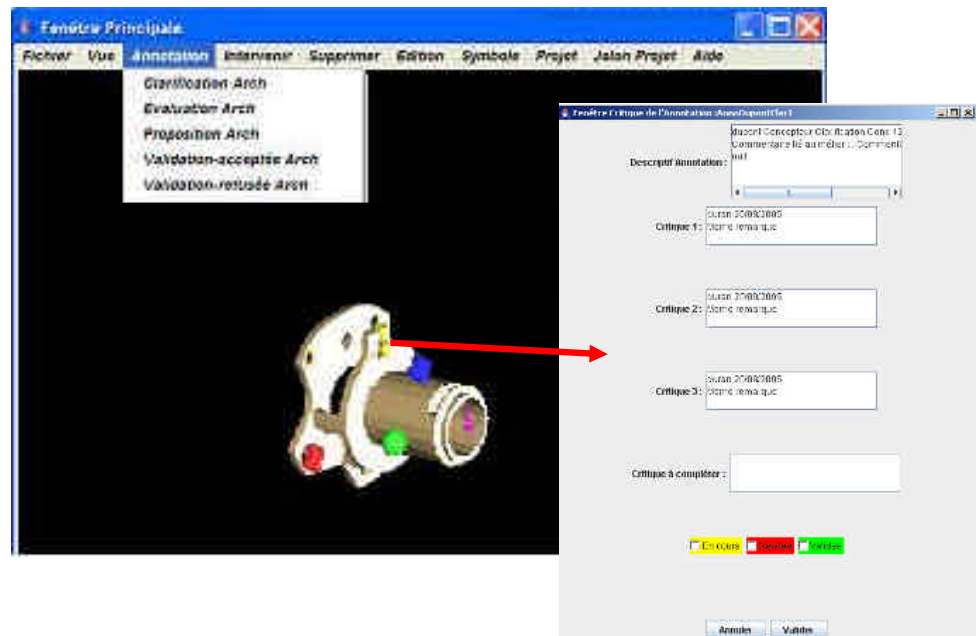

*Figure 3: Example of 3D annotations and a textbox attached to an annotation*

These annotation specifications offer the following benefits to the design communication:
- they allow the participants of a design team to annotate a 3D CAD instance,
- they provide an annotation structure (based on SAT) which allows participants to express domain-specific information with the intention to expressing it, and therefore achieving a conversational dimension,
- they allow an indexation structure to the participants, facilitating the annotation retrieval and re-use,
- They offer a way to record design rational, thanks to the illocutionary annotation acts.

## 7. CONCLUSION

In this paper, we described a SAT-based approach to support design communication with semantic annotations. Our approach leads to an annotation model in order for the design participants to express specific information with a particular intention in the design context where the annotation is created. This model offers a semantic annotation structure in order for the participants to express more complex information and improve the discussions around CAD models.

The next step of our research is to develop a technical structure in the form of an annotation server accessible through the web, which will allow the user to annotate 3D documents, and develop our approach on a greater number of design cases. At the same time, we are working on the improvement of the generic annotation symbols, and specify the particular locutionary and illocutionary annotation acts that could be used in an engineering design context.

Contact: Onur Hisarciklilar
G-SCOP Laboratory, INPG - Grenoble University
46 Avenue Félix Viallet
38000 cedex Grenoble
France
Phone. (+33) 4 56 52 89 06
Fax. (+33) 4 76 57 46 95
e-mail. Onur.Hisarciklilar@g-scop.inpg.fr